# Single monolayer ferromagnetic perovskite SrRuO$_3$ with high conductivity and strong ferromagnetism

*Yuki K. Wakabayashi,[1,*] Masaki Kobayashi,[2,3] Yoshiharu Krockenberger,[1] Takahito Takeda,[4] Kohei Yamagami,[5] Hideki Yamamoto,[1] and Yoshitaka Taniyasu[1]*

[1]NTT Basic Research Laboratories, NTT Corporation, Atsugi, Kanagawa 243-0198, Japan

[2]Center for Spintronics Research Network, The University of Tokyo, 7-3-1 Hongo, Bunkyo-ku, Tokyo 113-8656, Japan

[3]Department of Electrical Engineering and Information Systems, The University of Tokyo, Bunkyo, Tokyo 113-8656, Japan

[4]Graduate School of Advanced Science and Engineering, Hiroshima University, 1-3-1 Kagamiyama, Higashi-Hiroshima, 739-8526, Japan

[5]Japan Synchrotron Radiation Research Institute (JASRI), 1-1-1 Kouto, Sayo, Hyogo, 679-5198, Japan

[*]Corresponding author: yuuki.wakabayashi@ntt.com

Achieving robust ferromagnetism and high conductivity in atomically thin oxide materials is critical for advancing spintronic technologies. Here, we report the growth of a highly conductive and ferromagnetic single monolayer SrRuO$_3$ (SRO) having a high Curie temperature of ~154 K on DyScO$_3$ (110) substrates. The SrTiO$_3$ capping layer effectively suppresses surface reactions, which typically hinder ferromagnetism in atomically thin films. X-ray absorption spectroscopy and X-ray magnetic circular dichroism measurements revealed strong orbital hybridization between Ru 4$d$ and O 2$p$ orbitals in the SRO monolayer, which contributes to enhancement of the conductivity and ferromagnetic ordering of both the Ru 4$d$ and O 2$p$ orbitals. The resistivity of the single monolayer SRO on the better lattice-matched DyScO$_3$ substrate is approximately one-third of that of previously reported single monolayer SRO grown on an SrTiO$_3$ substrate. This study highlights the potential of monolayer SRO as a platform for two-dimensional magnetic oxide systems, offering new opportunities for the exploration of spintronic devices and quantum transport phenomena.

Since the discovery of two-dimensional ferromagnetism in van der Waals materials without dangling bonds on their surfaces,[1,2,3,4,5] significant interest has been drawn to their potential for combining the advantages of two-dimensional electron systems with those of magnetic materials, such as tunable state control by electric and magnetic fields.[6] However, practical applications of van der Waals ferromagnetic materials have been hindered by their still limited area size, which hampers scalable device fabrication and possible spintronics applications.[7,8] In contrast, reducing conventional three-dimensional ferromagnetic materials to a monolayer by controlling the film thickness encounters uncontrolled surface reactions with water, hydrogen, or carbon, destabilizing ferromagnetism in ambient conditions.[9,10,11,12,13,14]

The itinerant $4d$ ferromagnetic perovskite ruthenate $SrRuO_3$ (SRO) [bulk Curie temperature ($T_C$) = 165 K] is widely used as an epitaxial conducting layer in oxide electronics and spintronics owing to its many interesting behaviors in bulk and thin-film forms.[15-24] As a rare example of maintaining ferromagnetism in an atomically thin three-dimensional ferromagnet, a 1 monolayer (ML) SRO capped with $SrTiO_3$ (STO) shows domain magnetoresistance (MR),[25] indicating the presence of ferromagnetic domains. The STO cap effectively suppresses surface reactions, enabling the preservation of ferromagnetism even at 1 ML.[26] However, previous studies on 1 ML SRO detected only a small magnetic flux corresponding to 0.001–0.01 $\mu_B$/Ru, suggesting that the system is inhomogeneous, where ferromagnetic domains coexist with non-magnetic regions.[25] Additionally, its $T_C$ of 25 K is significantly lower than that of the bulk, and room-temperature conductivity is approximately one order of magnitude lower than that of bulk and thick-film SRO.[25] Furthermore, the electronic states of 1 ML conductive SRO remain unexplored,[27,28] and a deeper understanding of its intrinsic electronic properties is essential. Therefore, achieving 1 ML ferromagnetic SRO with enhanced $T_C$, magnetization, and conductivity, as well as clarifying its electronic structure, is crucial for its potential spintronics applications. In addition, the recent observations of the quantum transport of Weyl fermions[29,30,31,32] with high mobility exceeding 10,000 cm$^2$/Vs and linear band dispersions of Weyl cones[33] in SRO make the realization of highly conductive 1 ML ferromagnetic SRO desirable for achieving a two-dimensional magnetic Weyl semimetal state and exploring novel quantum transport phenomena.

In this study, we grew highly conductive ferromagnetic 1 ML SRO having a high $T_C$ of 154 K capped with STO on a $DyScO_3$ (DSO) (110) substrate using machine-learning-assisted molecular beam epitaxy (ML-MBE).[34,35] The electronic and magnetic states of the fabricated 1 ML SRO were investigated through X-ray absorption spectroscopy (XAS) and X-ray magnetic circular dichroism (XMCD), which are highly sensitive to the local electronic structure and element-specific magnetic properties.[36-39] The resistivity of the single monolayer SRO on the better lattice-matched $DyScO_3$ substrate is approximately one-third of that of previously reported 1 ML SRO capped with STO on a STO substrate.[25] The XAS measurements revealed strong hybridization between Ru $4d$ and O $2p$ orbitals, which played a crucial role in achieving the high conductivity of the 1 ML SRO. The XMCD measurements indicated the presence of spontaneous magnetization in both Ru $4d$ and O $2p$ orbitals, demonstrating the ferromagnetic ordering in 1 ML SRO. The magnetic moment of oxygen was induced by charge transfer between the Ru $4d$ and O $2p$ orbitals. The magnetic moment of Ru reached 0.19 $\mu_B$/Ru at 14 K and 1.92 T, which is about 30% of that observed in thick SRO films.

We grew 1 ML epitaxial SRO capped with STO on DSO(110) substrate [Fig. 1(a)] by ML-MBE, in which the growth conditions were optimized by Bayesian optimization, a machine learning technique for parameter optimization.[34,35] The growth conditions are the same as those described in Ref. [40]. Further information about the MBE setup is described elsewhere.[19,41] For comparison, a thick 25 ML SRO film capped with STO was also grown on a DSO substrate at the same growth conditions. The film thicknesses were controlled from the growth rate of 1.1 Å/s deduced from the periods of the Laue fringes in $\theta-2\theta$ x-ray diffraction patterns of a 60-nm thick SRO film.[40] Since the lattice mismatch between bulk SRO (3.93 Å) and DSO substrate (3.944 Å) is smaller than that between bulk SRO and STO substrate (3.905 Å), fewer defects are expected at the interface.[18,40] The surface morphology of the 1 ML SRO and STO cap layers were observed by atomic force microscopy. The 1 ML SRO surface shows a step and terrace structure with a height of ~0.4 nm [Fig. 1(b)], corresponding to the single unit cell. Although the step-and-terrace structure is no longer visible on the STO-capped surface [Fig. 1(c)], the root mean square of the surface roughness remains as small as 0.14 nm, indicating that a flat surface is maintained. The XAS and XMCD measurements were performed at the helical undulator beamline BL25SU of SPring-8.[42] For the XMCD measurements, absorption spectra for circularly polarized X rays with the photon helicity parallel ($\mu^+$) or antiparallel ($\mu^-$) to the spin polarization were recorded in the total-electron-yield (TEY) mode. The measurement temperature was 14 K. The angle of the external magnetic fields from the sample surfaces was set to $\theta = 90°$. The incident X rays were tilted 10° to the magnetic field.[43]

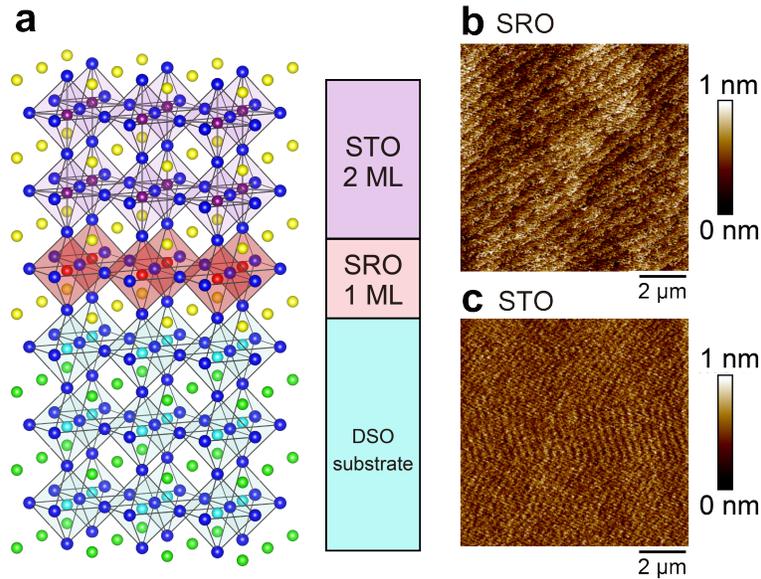

Figure 1. (a) Schematic diagrams of the sample and crystal structures of the 1 ML SRO capped with STO on DSO substrate. In the schematic of the crystal structure, yellow, blue, purple, red, green, and cyan spheres indicate strontium, oxygen, titanium, ruthenium, dysprosium, and scandium, respectively. (b), (c) AFM images of the surface of the (b) 1 ML SRO and (c) STO cap layers.

Figure 2 shows the temperature dependence of resistivity for the 1 ML and 25 ML SRO/DSO films. Both films show a kink (arrows in Fig. 2), at which the ferromagnetic transition occurs and spin-dependent scattering is suppressed.[18] Note that, generally, $T_C$ as the kink position is a few kelvins lower than the values measured from the temperature dependence of the magnetization.[44] The $T_C$ for 1 ML SRO on DSO (~154 K) is much higher than that for 1 ML SRO on STO (~25 K).[25] The $T_C$ value is comparable to that at 25 ML (~164 K), and it is evident that ferromagnetic order is maintained in 1 ML SRO.[25] The resistivity of 1 ML SRO continues to decrease even below $T_C$, reaching a minimum at 71 K before starting to increase, possibly owing to the localization of the charge carriers. The resistivity of 1 ML SRO has a value between 253 and 339 $\mu\Omega$·cm. This resistivity is larger than that of bulk SRO or 25 ML SRO; however, it is significantly small, being about one-third of the lowest reported value for 1 ML SRO on STO substrates in previous studies.[25,45] The high $T_C$ and conductivity of 1 ML SRO on DSO are attributed to the smaller lattice mismatch with the DSO substrate than that with the STO substrate, resulting in fewer defects in our high-quality 1 ML SRO.[40]

The ferromagnetic behavior of 1 ML SRO is prominently reflected in its magnetotransport properties. Figure 2(b) shows the change in the MR $[\rho(B) - \rho(0\,\text{T})]/\rho(0\,\text{T})]$ of the 1 ML SRO across the $T_C$ (~154 K). Below 130 K, the anisotropic magnetoresistance (AMR),[46] which is proportional to the relative angle between the electric current and the magnetization, exhibits a clear ferromagnetic hysteresis. Even at 140 K, while the hysteresis disappears, a non-parabolic MR indicative of the presence of ferromagnetic order in SRO is still observed.[25] We observed a continuous change into a parabolic MR, characteristic of non-magnetic conductors, across the $T_C$ from 140 K to 160 K.

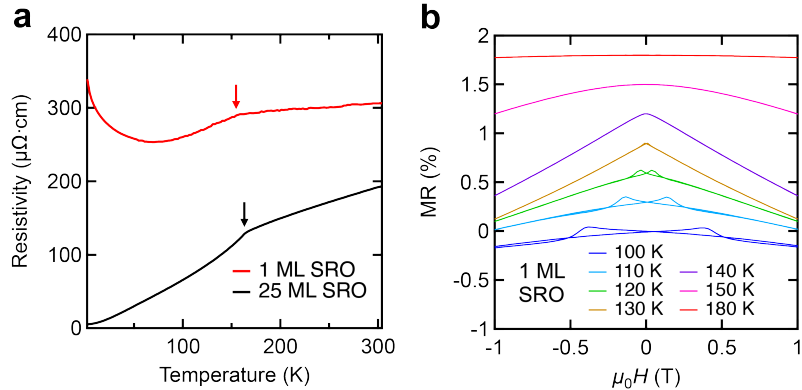

Figure 2. (a) Temperature dependence of resistivity for 1 ML and 25 ML SRO films on DSO substrates. Arrows indicate kinks. (b) MR of the 1 ML SRO with $\mu_0 H$ applied in the out-of-plane [001] direction of the SrTiO$_3$ substrate. In (b), arrows indicate the direction of the magnetic field sweep.

Figure 3(a) shows the Ru $M_{2,3}$-edge XAS spectra for the 1 ML SRO/DSO (black line) and the STO substrate (red dashed line) at 14 K. Although the Ru $M_3$-edge overlaps the Ti $L_{2,3}$ peaks from the STO cap layer, a distinct Ru $M_2$-edge peak is observed for the 1 ML SRO. To investigate the unoccupied electronic states hybridized with the O $2p$ orbitals and the magnetic properties of the O ions, we also measured the O $K$-edge XAS spectra [Fig. 3(b)]. The absorption peak at 529 eV corresponds to the transitions to the coherent Ru $4d$ $t_{2g}$ states hybridized with the O $2p$ states.[47,48] Transitions to the incoherent part of the Ru $4d$ $t_{2g}$ states are also observed in the energy range of 530–532 eV,[47,49] accompanied by contributions from the Ti $3d$ $t_{2g}$ states of the STO cap layer at 531.1 eV. The intense coherent Ru $4d\,t_{2g}$ peak indicates the long lifetimes of quasiparticles in the hybridized Ru $4d\,t_{2g}$-O $2p$ states,[48,50] consistent with the high conductivity in the 1 ML SRO. This contrasts with the 1 ML SRO grown on the STO substrate by pulsed laser deposition without the STO cap layer.[50] Without the STO cap layer, orbital hybridization weakens, the coherent peak disappears, and insulating transport characteristics are observed.[50] This result indicates that the STO cap reduces surface disorder and scattering, thereby preserving the ferromagnetic metallic state in the 1 ML SRO. The magnetic moments of Ru and O, as well as their ferromagnetic ordering, arising from the coherent $4d\,t_{2g}$-O $2p$ hybridized states, were observed using XMCD. Figure 3(c) shows Ru $M_2$-edge XMCD spectra for the 1 ML SRO at 14 K under a magnetic field $\mu_0H$ of 0, 0.49, 0.97, and 1.92 T. Although the XMCD intensity decreases as the magnetic field decreases, an XMCD signal due to spontaneous magnetization is still observed at 0 T, providing clear evidence of ferromagnetic ordering. It is known that the orbital magnetic moment of oxygen $2p$ orbitals is proportional to the $K$-edge XMCD intensity.[36,51,52,53] A substantial O $K$ XMCD signal is observed in the energy range corresponding to the coherent Ru $4d\,t_{2g}$ peak [Fig. 3(d)]. Similarly, with the Ru magnetic moment, a substantial orbital magnetic moment of the O $2p$ states, which arises from charge transfer to the Ru $4d\,t_{2g}$ states via orbital hybridization,[50,54] decreases with decreasing magnetic field, while a finite spontaneous XMCD signal remains. It should be noted here that the line shapes of the Ru $M_2$ and O $K$ XMCD spectra are like those in a thick SRO film,[47,49,54] indicating that the Ru $4d\,t_{2g}$ orbital well hybridized with the O $2p$ one even in the film with the thickness of 1 ML.

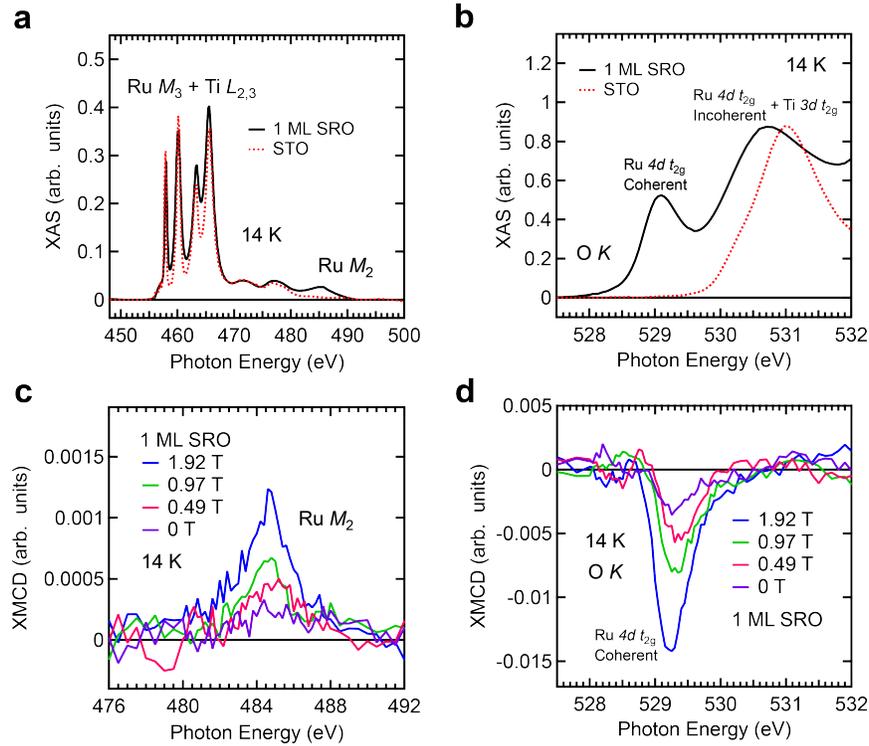

Figure 3. (a) Ru $M_{2,3}$-edge and (b) O $K$-edge XAS spectra for the 1 ML SRO/DSO (black line) and the STO substrate (red dashed line) at 14 K. (c) Ru $M_2$-edge and (d) O $K$-edge XMCD spectra for the 1 ML SRO at 14 K under a magnetic field $\mu_0H$ of 0, 0.49, 0.97, and 1.92 T.

In XMCD measurements, the magnetic moments of each element are proportional to the ratios of XMCD to XAS integrated intensities of the particular absorption edges.[36] According to the XMCD sum rules, the orbital magnetic moment $m_{orb}$ and the spin magnetic moment $m_{spin}$ of the $Ru^{4+}$ $4d$ states are described as follows:[36,37,38]

$$m_{orb} = -\frac{4(10-n_{4d})}{3r} \int_{M_2+M_3}^{\square} (\mu^+ - \mu^-) dE, \qquad (1)$$

$$m_{spin} + m_T = -\frac{2(10-n_{4d})}{r} [\int_{M_3}^{\square} (\mu^+ - \mu^-) dE - 2 \int_{M_2}^{\square} (\mu^+ - \mu^-) dE]. \qquad (2)$$

Here, $r = \int_{M_2+M_3}^{\square} (\mu^+ + \mu^-) dE$, and $n_{4d}$ is the number of electrons in $4d$ orbitals, which is assumed to be four. For ions in octahedral symmetry, the magnetic dipole moment $m_T$ is a small number and can be neglected compared to $m_{spin}$.[55] As shown in these equations, deriving the magnetic moment of Ru typically requires measuring both $\mu^+$ and $\mu^-$ spectra at the $M_3$ and $M_2$ edges. However, in SRO, the absorption spectral shapes at the $M_3$ and $M_2$ edges are known not to change, regardless of film thickness, epitaxial strain, electrical conductivity, or the direction of the applied magnetic field.[47,49,54,56] Therefore, we can determine the total magnetic moment of Ru ($m_{Ru}$ = $m_{spin}$ + $m_{orb}$) by integrating only $\mu^+$ and $\mu^-$ spectra at the $M_2$ edge from 471 to 495 eV. Similarly, the orbital magnetic moment of oxygen $m_{orb}^O$ was evaluated by integrating the O $K$-edge XMCD/XAS intensity from 527 to 531 eV, where finite O $K$-edge XMCD intensity appears [Fig. 2(d)].[54] Although it is difficult to quantitatively determine the orbital magnetic moments of the SRO films from this sum rule since the number of O $2p$ holes is not known, the ratio of the orbital magnetic moment in the same sample depending on the applied magnetic field can be determined quantitatively.

The magnetic field $\mu_0 H$ dependence of $m_{Ru}$ and $m_{orb}^O$ for the 1 ML SRO/DSO, estimated from the XAS and XMCD spectra in Fig. 3, at 14 K are plotted in Fig. 4. Here, the $m_{orb}^O$ at 1 T is scaled to 3/7 of the $m_{Ru}$ at 1 T, as it is known that oxygen contributes 30% of the total magnetization in bulk $SrRuO_3$.[57] The $m_{Ru}$ value at 1.92 T is 0.19 $\mu_B$/Ru, which is approximately 30% of the value (0.51 $\mu_B$/Ru) observed in 60 nm (150 ML) SRO.[56] The spontaneous magnetization of $m_{Ru}$, which is intrinsically important for spintronics applications of 1 ML SRO, is 0.06 $\mu_B$/Ru. The magnetic field dependence of $m_{orb}^O$ agrees with that of $m_{Ru}$, indicating a strong magnetic coupling between Ru and O. This magnetic coupling is considered to arise from the strong orbital hybridization between Ru $4d$ and O $2p$ states, which has been confirmed in the O $K$-edge XAS spectra as described earlier.

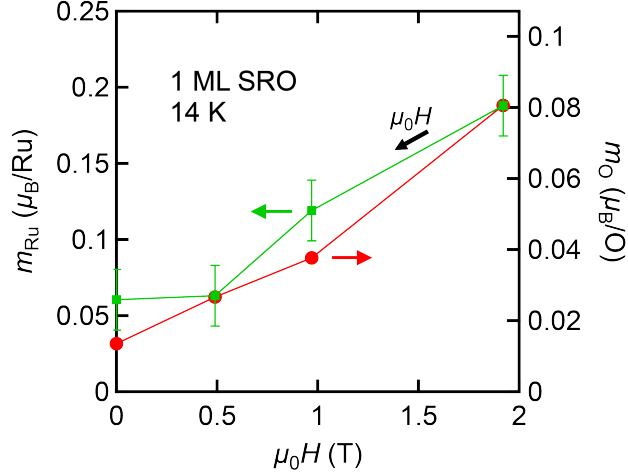

Figure 4. Magnetic field $\mu_0 H$ dependence of the total magnetic moment of Ru (green squares) and the orbital magnetic moment of O (red circles) for the 1 ML SRO/DSO at 14 K. The black arrow indicates the direction of the magnetic field sweep.

In summary, we have demonstrated the successful growth of highly conductive and ferromagnetic 1 ML SRO having the $T_C$ of ~154 K capped with STO using ML-MBE. Our XAS and XMCD measurements revealed that the strong orbital hybridization between Ru $4d$ and O $2p$ orbitals plays a crucial role in achieving high conductivity and ferromagnetic ordering in 1 ML SRO. The Ru magnetic moment of 0.19 $\mu_B$/Ru at 1.92 T and 14 K, as well as a finite spontaneous magnetization of 0.06 $\mu_B$/Ru, was observed. The high-quality 1 ML SRO exhibited a resistivity of approximately one-third of that reported for previously studied 1 ML SRO films, which implies reduced defect density due to the small lattice mismatch between SRO and the DSO substrate. The observed spontaneous magnetization and enhanced conductivity indicate the potential of 1 ML SRO for future spintronics applications. These results open up new possibilities for the development of scalable two-dimensional ferromagnetic oxide heterostructures for spintronics and quantum devices.

## AUTHOR INFORMATION

### Corresponding Author

Yuki K. Wakabayashi ─ *NTT Basic Research Laboratories, NTT Corporation, Atsugi, Kanagawa 243-0198, Japan*; Email: yuuki.wakabayashi@ntt.com

ACKNOWLEDGMENT

This work was partially supported by the Spintronics Research Network of Japan (Spin-RNJ). The synchrotron radiation experiments were performed at the BL25SU of SPring-8 with the approval of the Japan Synchrotron Radiation Research Institute (JASRI) (Proposals No. 2024B1409).